%% file: Lattice2017_DpiK_SALERNO.tex
\documentclass[epj]{webofc}
\usepackage[utf8]{inputenc}
\usepackage[varg]{txfonts}   
\usepackage{booktabs}
\usepackage{xcolor}
\definecolor{darkred}{rgb}{0.4,0.0,0.0}
\definecolor{darkgreen}{rgb}{0.0,0.4,0.0}
\definecolor{darkblue}{rgb}{0.0,0.0,0.4}
\usepackage[bookmarks,linktocpage,colorlinks,
linkcolor = darkred,
urlcolor  = darkblue,
citecolor = darkgreen]{hyperref}

\usepackage{array} 

\usepackage{amsmath,amssymb,mathrsfs,textcomp,amscd,bbm} 

\usepackage{graphicx} 
\DeclareGraphicsExtensions{.pdf,.png,.jpg,.mps,.ps,}
\graphicspath{{Immagini_Dpi17/}}

\usepackage{braket}

\usepackage{subfigure,sidecap}
\wocname{EPJ Web of Conferences}
\woctitle{Lattice2017}
\input{definitions.tex}

\begin{document}
%
\selectlanguage{english}
\title{%
	\boldmath$D\to\pi$ and \boldmath$D\to K$ semileptonic form factors with \boldmath$N_f=2+1+1$ twisted mass fermions\thanks{Presented at Lattice 2017 - 35th International Symposium on Lattice Field Theory, 18-24 June 2017, Granada.}
}
\author{%
	\firstname{Vittorio} \lastname{Lubicz}\inst{1,2} \and
	\firstname{Lorenzo} \lastname{Riggio}\inst{2} \and
	\firstname{Giorgio} \lastname{Salerno}\inst{1,2} \and	
	\firstname{Silvano} \lastname{Simula}\inst{2} \and
		\firstname{Cecilia} \lastname{Tarantino}\inst{1,2} 
}
\institute{%
	Dipartimento di Matematica e Fisica, Università Roma Tre, Rome, Italy.
	\and
	Istituto Nazionale di Fisica Nucleare (INFN), Sezione di Roma Tre, Rome, Italy.
}
\abstract{%
We present a lattice determination of the vector and scalar form factors of the $D\to\pi(K)\ell\nu$ semileptonic decays, which are relevant for the extraction of the CKM matrix elements $|V_{cd}|$ and $|V_{cs}|$ from experimental data. Our analysis is based on the gauge configurations produced by the European Twisted Mass Collaboration with $N_f=2+1+1$ flavors of dynamical quarks. We simulated at three different values of the lattice spacing and with pion masses as small as $210\MeV$. The matrix elements of both vector and scalar currents are determined for a plenty of kinematical conditions in which parent and child mesons are either moving or at rest. Lorentz symmetry breaking due to hypercubic effects is clearly observed in the data and included in the decomposition of the current matrix elements in terms of additional form factors. After the extrapolations to the physical pion mass and to the continuum limit the vector and scalar form factors are determined in the whole kinematical region from $q^2 = 0$ up to $q^2_{\rm max} = (M_D - M_{\pi(K)})^2$ accessible in the experiments, obtaining a good overall agreement with experiments, except in the region at high values of $q^2$ where some deviations are visible.
}
\maketitle

\section{Introduction and simulation details}

Flavor physics, the branch of particle physics that studies transitions between different quarks and leptons, plays a fundamental role both for an indirect search of New Physics (NP) and also to put stringent constraint on the Standard Model (SM). Unlike the gauge sector, which is totally fixed by the symmetry $SU(3)_C\otimes SU(2)_L\otimes U(1)_Y$, the flavor sector is completely loose and is characterized by masses and quark mixing that are free parameters. Moreover, because of its highly non trivial structure, it is particularly sensitive to many NP scenarios.

In this contribution we present the first $N_f = 2+1+1$ LQCD calculation of the vector and scalar form factors $f_+^{D \pi(K)}(q^2)$ and $f_0^{D \pi(K)}(q^2)$, governing the semileptonic $D \to \pi(K) \ell \nu$ decays, relevant for the extraction of the CKM matrix elements $|V_{cd}|$ and $|V_{cs}|$~\cite{Lubicz:2017syv}. The analysis has been carried out using the gauge configurations generated by the European Twisted Mass Collaboration (ETMC) with $N_f = 2 + 1 + 1$ dynamical quarks, which include in the sea, besides two light mass-degenerate quarks, also the strange and the charm quarks \cite{Baron:2010bv,Baron:2011sf}. In order to take under control discretization and finite volume effects, simulations have been performed at various lattice volumes and using three values of the lattice spacing in the range $0.06 - 0.09\, \fm$, with  pion masses from $\approx 210\MeV$ up to $\approx 450\MeV$.
Sea and light valence quarks were simulated using the Wilson Twisted Mass Action \cite{Frezzotti:2003xj,Frezzotti:2003ni}, while the valence charm quark was implemented through the Osterwalder-Seiler action \cite{Osterwalder:1977pc}. Gauge fields were simulated using the Iwasaki gluon action \cite{Iwasaki:1985we}. Such a setup guarantees for an automatic $\mathcal{O}(a)$ improvement as the Twisted Mass Action for light and see quarks has been taken at maximal twist \cite{Frezzotti:2003ni,Frezzotti:2004wz}. We refer the reader to Ref.~\cite{Carrasco:2014cwa} for more details on the lattice setup.

Non-periodic boundary conditions for the quark fields have been used in order to inject momenta on the lattice \cite{Bedaque:2004kc,deDivitiis:2004kq,Guadagnoli:2005be} - with values in the range $150 - 650\MeV$ - obtaining a plenty of kinematical conditions in which parent and child mesons are either moving or at rest.
The lattice data exhibit a remarkable breaking of the Lorentz symmetry~\cite{Lubicz:2017syv} (see also preliminary results in Refs.~\cite{Carrasco:2015bhi,Lubicz:2016wwx}) due to hypercubic effects both for the $D\to \pi$ and the $D\to K$ channels.
Hypercubic artefacts appear to be driven by the difference between the parent and the child meson masses, which may represent an important warning in the case of semileptonic $B$-meson decays.
We present the subtraction of such hypercubic effects and the determination of the physical, Lorentz-invariant, semileptonic vector and scalar form factors in the whole experimentally accessible range in $q^2$, i.e.~from $q^2 = 0$ up to $q^2_{\rm max} = (M_D - M_{\pi(K)})^2$, at variance with respect to existing LQCD calculations (see Ref.~\cite{Aoki:2016frl}), which provide only the value of the vector form factor at zero 4-momentum transfer.

\section{Vector and scalar form factors}

Let us introduce the local bare currents $V_\mu = \bar{c} \gamma_\mu q$ and $S = \bar{c} q$, where $q = d (s)$. In our lattice setup we employ maximally twisted fermions and thus the vector and scalar currents renormalize multiplicatively \cite{Frezzotti:2003ni}, i.e. $\widehat{V}_\mu =  {\cal{Z}}_V\cdot V_\mu$ and $\widehat{S}=  {\cal{Z}}_P\cdot S$, where $\widehat{V}_\mu$ and $\widehat{S}$ are the renormalized vector and scalar densities, with ${\cal Z}_V$ and ${\cal Z}_P$ being the corresponding renormalization constants (RCs).

The matrix elements $\braket{P(p_P) | \widehat{V}_\mu | D(p_D)}\equiv\braket{\widehat{V}_\mu}$ and $\braket{P(p_P) | {S} | D(p_D)}\equiv\braket{{S}}$, as required by the Lorentz symmetry, can be decomposed into the vector and scalar form factors $f_+(q^2)$ and $f_0(q^2)$:
\begin{eqnarray}
\label{eq:vectorL}
\braket{\widehat{V}_\mu}&\!\!\!=\!\!\!&\left[p_{D\mu}+p_{P\mu}-q_\mu\,(m^2_D-m^2_{P})/q^2\right]\, f_+(q^2) + q_\mu\, f_0(q^2) \,(m^2_D-m^2_P)/q^2+ {\cal{O}}(a^2)\,,\\[2mm]
\label{eq:scalarL}
\braket{{S}} & \!\!\!=\!\!\! & f_0(q^2)\,(M_D^2 - M_P^2)/(\mu_c - \mu_q)  + {\cal{O}}(a^2)\,,
\end{eqnarray}
where $q=p_D-p_P$ is the four-momentum of the outgoing lepton pair, $P$ stands either for a $\pi$ or a $K$ meson, and $\mu_{c(q)}$ is the charm (light) bare quark mass. The vector and scalar matrix elements have been extracted from the large time distance behavior of five ratios, $R_\mu$ $(\mu=1,2,3,4)$ and $R_S$, which are given by:
\begin{eqnarray}
\label{eq:Rmu} 
R_\mu(t,\vec{p}_D, \vec{p}_P) \!\!\!&\equiv&\!\!\! 4\, p_{D \mu}\, p_{P \mu}\, \frac{C^{DP}_{V_\mu}(t, t^\prime, \vec{p}_D, \vec{p}_P) \, 
	C^{P D}_{V_\mu}(t, t^\prime, \vec{p}_P, \vec{p}_D)} {C^{P P}_{V_\mu}(t, t^\prime, \vec{p}_P, \vec{p}_P) \, 
	C^{DD}_{V_\mu}(t, t^\prime, \vec{p}_D, \vec{p}_D)}~ _{\overrightarrow{t\gg a\, \, (t^\prime-t)\gg a}} ~\lvert\braket{\widehat{V}_\mu} \rvert^2 \,,\\[2mm]
\label{eq:RS}
R_S(t,\vec{p}_D, \vec{p}_P) \!\!\!&\equiv&\!\!\! 4\, E_D\, E_P\,\frac{C^{DP}_S(t, t^\prime,\vec{p}_D, \vec{p}_P)\, C^{P D}_S(t, t^\prime,\vec{p}_P, \vec{p}_D)} {\widetilde{C}_2^D \left(t^\prime,\vec{p}_{D} \right)\, \widetilde{C}_2^P \left(t^\prime,\vec{p}_{P} \right)}~ _{\overrightarrow{t\gg a\, \, (t^\prime-t)\gg a}}~\lvert\braket{S} \rvert^2  \,.
\end{eqnarray}
In Eqs.~(\ref{eq:Rmu},\ref{eq:RS}) $C^{DP}_{\Gamma}$ ($\Gamma=V_\mu,\,S$) and $\widetilde{C}_2^M$ are respectively the 3-point correlation function between the $D$ and the $P$ mesons and a ``modified'' 2-point correlation function for the meson $M$ in which the backward signal is cancelled (see Eq.~(17) of Ref.~\cite{Lubicz:2017syv}). At large time distances they are defined as:
\begin{eqnarray}
\label{eq:C3_larget}        
&&\qquad C^{DP}_{\widehat{\Gamma}}\left(  t,\, t^\prime,\, \vec{p}_D,\, \vec{p}_P \right) ~ _{\overrightarrow{t\gg a\,, \, (t^\prime-t)\gg a}} ~ \frac{Z_P Z_D^*}{4E_P E_D}\, \braket{P(p_P)|\widehat{\Gamma}|D(p_D)}\, e^{-E_D t}\, e^{-E_P (t^\prime - t)}~,\\[2mm]
\label{eq:C2_tilde_larget}
&&\qquad \widetilde{C}_2^{D(P)} \left(t,\,\vec{p}_{D(P)}\right) \quad~ _{\overrightarrow{t \gg a}} ~ \frac{Z_{D(P)}}{2\, E_{D(P)}}\,e^{-E_{D(P)} t} ~ ,
\end{eqnarray}
where $E_{D}$ and $E_{P}$ are the energies of the $D$ and $P$ mesons, while $Z_D$ and $Z_P$ are the matrix elements $\braket{0\lvert\,P_5^D(0)\,\rvert\,D(\vec{p}_D) }$ and $\braket{0\lvert\,P_5^P(0)\,\rvert\,P(\vec{p}_{P})}$, which depend on the meson momenta $\vec{p}_D$ and $\vec{p}_P$ because of the use of smeared interpolating fields.
From the 2- and 3-point correlators we are able to extract the matrix elements $\braket{\widehat{V}_\mu}$ and $\braket{S}$, which allow us to determine $f_+(q^2)$ and $f_0(q^2)$ as the best-fit values of Eqs.~(\ref{eq:vectorL}) and (\ref{eq:scalarL}). This procedure, after a small interpolation to the physical values of the strange and charm quark masses $m_s^{phys}$ and $m_c^{phys}$(determined in Ref.~\cite{Carrasco:2014cwa}), is applied to each choice of the parent and child meson momenta. 

\begin{figure}[htb]
	\centering
	\includegraphics[width=6.7cm,clip]{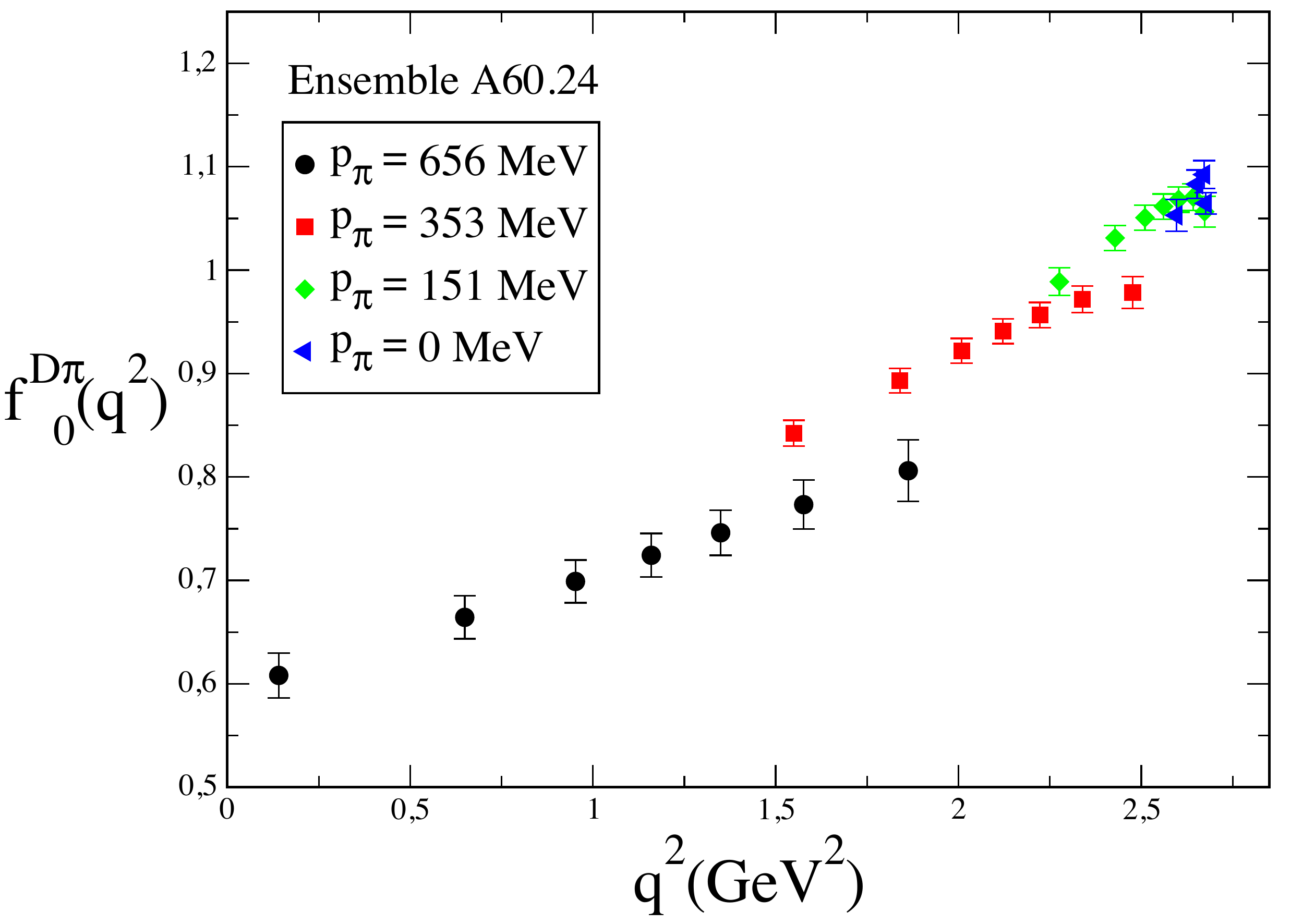}
	\includegraphics[width=6.7cm,clip]{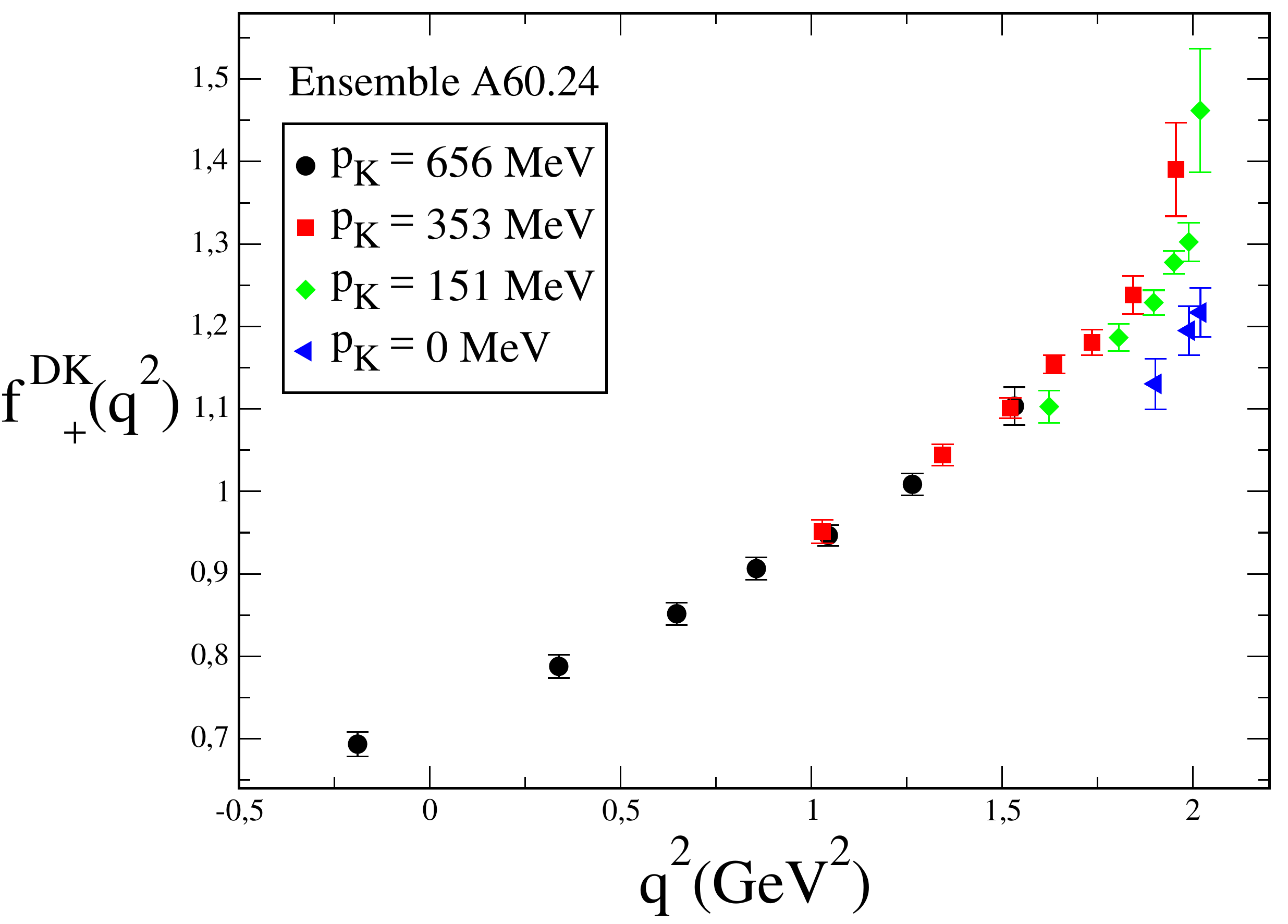}
	\caption{\it Momentum dependence of the semileptonic $D\to\pi$ scalar form factor (left panel) and $D\to K$ vector form factor (right panel), in the case of the gauge ensemble A60.24 corresponding to $M_\pi \approx 385$ MeV~\protect\cite{Lubicz:2017syv}. Different markers and colors distinguish different values of the child meson momentum.}
	\label{fishbone}
\end{figure}

An example of our results are illustrated in Fig.~\ref{fishbone}, where the momentum dependence of the scalar $f_0^{D\pi}(q^2)$ and vector $f_+^{DK}(q^2)$ form factors is reported for the ETMC ensemble A60.24.
It is clear how the Lorentz-covariant decomposition (\ref{eq:vectorL}-\ref{eq:scalarL}) is not adequate to describe the lattice data, in fact the extracted form factors, beyond statistical uncertainties, do not depend only on the squared 4-momentum transfer $q^2$ (and the parent and child meson masses), but also on the value of the child (or parent) meson momentum. The breaking of the Lorentz symmetry is mainly due to hypercubic discretization effects, which are of the order ${\cal{O}}(a^2)$ because of the ${\cal{O}}(a)$-improvement of our lattice setup. The possible contribution due to finite volume effects has been investigated by comparing results coming from two ensembles, which correspond to the same pion mass and lattice spacing, but have different lattice sizes. Hypercubic effects have been found to be compatible in the two cases, so FSEs cannot be the source of the observed behavior in the form factors.

\begin{figure}[htb]
	\centering
	\makebox[\textwidth][c]{\includegraphics[width=6.8cm,clip]{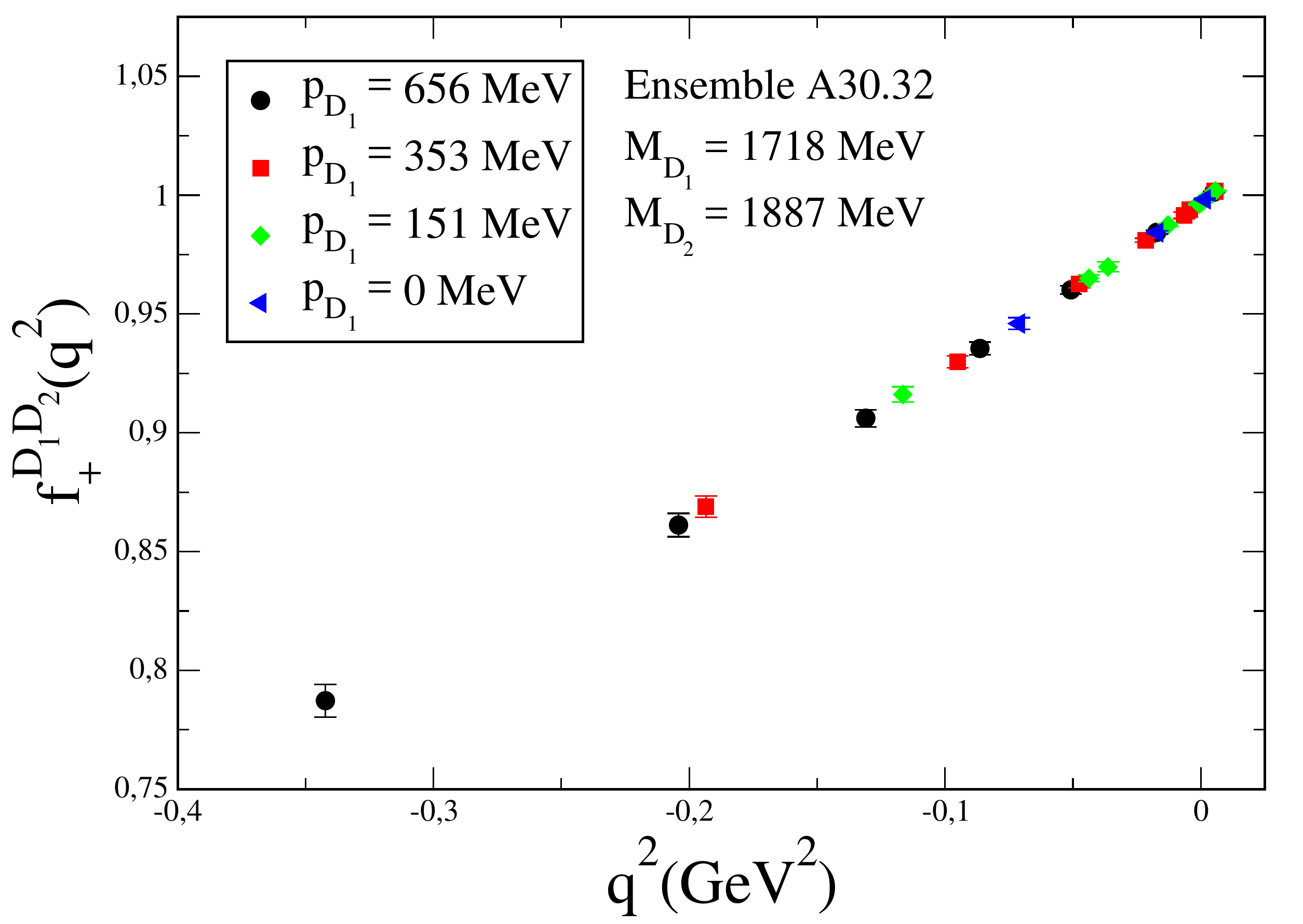}
		\includegraphics[width=6.7cm,clip]{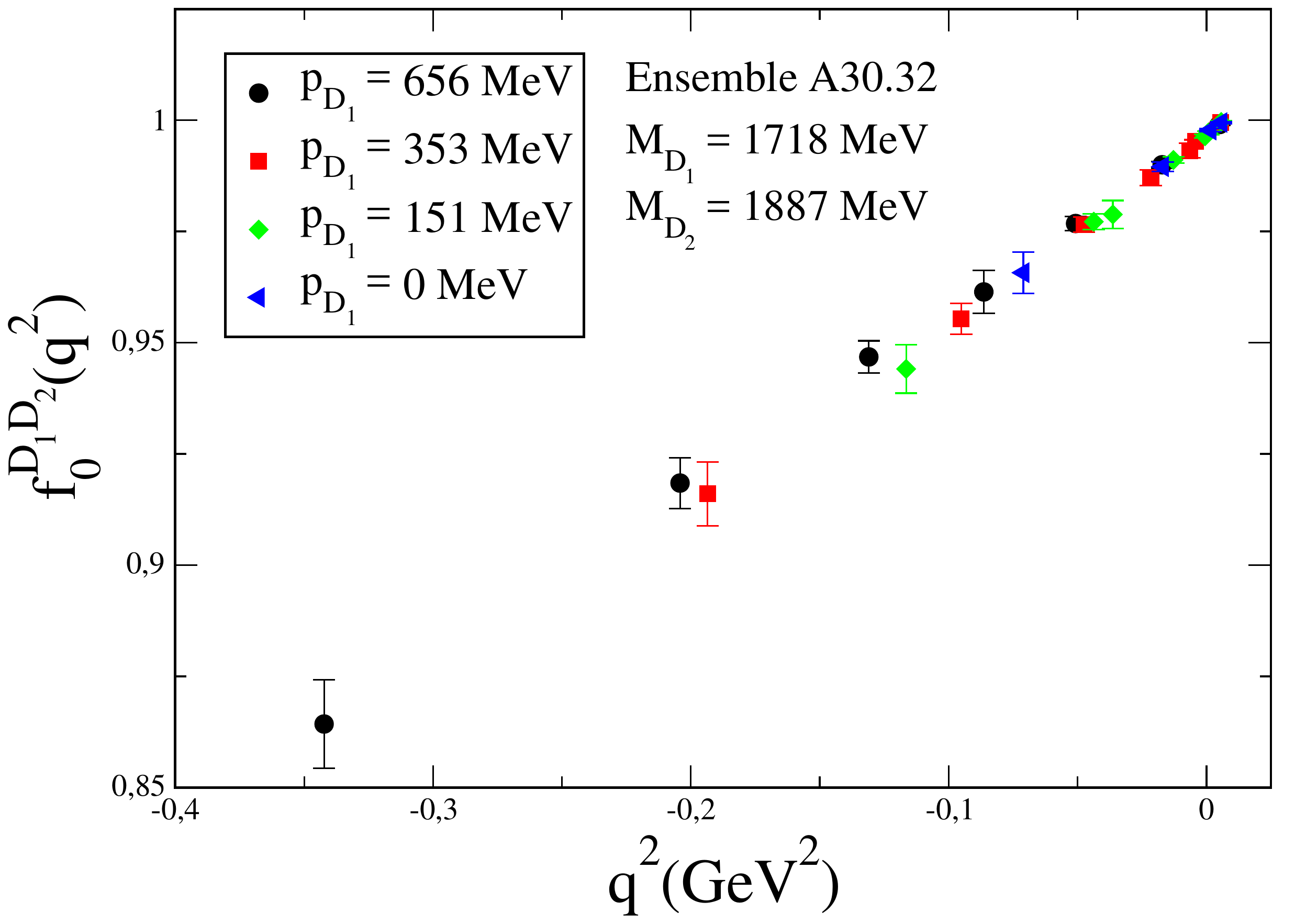}
	}
	\caption{\it Momentum dependence of the vector (left panel) and scalar (right panel) form factors relevant for the semileptonic decay between two charmed PS mesons, $D_1$ and $D_2$, with masses close to the $D$-meson one. The plot refers to the gauge ensemble A30.32~\protect\cite{Lubicz:2017syv}, in which $D_1$ and $D_2$ have masses equal to $1718$ MeV and $1887$ MeV, respectively. 
	}
	\label{fig:DtoD}
\end{figure}

A further important point to be stressed, is that no evidence of a Lorentz symmetry breaking has been observed in the behavior of the form factors relevant for the $K\to\pi\ell\nu$ decay \cite{Carrasco:2016kpy}.
A possible reason for this fact is that hypercubic artifacts may be mostly governed by the difference between the
parent and the child meson masses. 
Following this hypothesis, we studied the semileptonic transition between two charmed PS mesons with similar masses close to the $D$-meson one.
The results of such analysis are given in Fig.~\ref{fig:DtoD}, where the momentum dependence of the form factors, within statistical uncertainties, are clearly free of hypercubic effects and, as required by the Lorentz symmetry, are only function of $q^2$. This confirms the dependence of hypercubic effects upon the mass difference between the parent and the child mesons. 
However, further investigations are needed in order to understand better this important issue, which might become crucial in the case of semileptonic $B$-meson decays.

\section{Global fit}

As shown in the previous Section the form factors $f_+$ and $f_0$ for both the $D \to \pi$ and $D \to K$ decays exhibit a sizeable Lorentz-symmetry breaking due to hypercubic effects. 
A possible way to describe hypercubic artifacts is to address them directly on the vector and scalar matrix elements in terms of Euclidean momenta, given by $q_\mu^E = \left( \vec{q}, ~ q_4 \right) = \left( \vec{q}, \, -iq_0 \right)$. In particular, we considered the following decomposition of the vector and scalar currents:
\begin{eqnarray}
\centering
\label{eq:V_decomposition}
&&\qquad\qquad\qquad\qquad \braket{P(p_P) | \widehat{V}_\mu^E | D(p_D)} = \braket{\widehat{V}_\mu^E}_{\rm Lor} + \braket{\widehat{V}_\mu^E}_{\rm hyp} \,,\\[2mm]
\label{eq:S_decomposition}
&&\qquad\qquad\qquad\qquad \braket{P(p_P) | S | D(p_D)} = \braket{S}_{\rm Lor} + \braket{S}_{\rm hyp} \,,
\end{eqnarray}
in which $\braket{\widehat{V}_\mu^E}_{\rm Lor}$ and $\braket{S}_{\rm Lor}$ are the Lorentz-covariant terms defined in Eqs.~(\ref{eq:vectorL},\ref{eq:scalarL}), while $\braket{\widehat{V}_\mu^E}_{\rm hyp}$ and $\braket{S}_{\rm hyp}$ are hypercubic artifacts given by
\begin{eqnarray}
\label{eq:vector_hypercubic}
&&\braket{\widehat{V}_\mu^E}_{\rm hyp} = a^2 \left[ \left( q_\mu^E \right)^3 ~ H_1+ \left( q_\mu^E \right)^2 P_\mu^E ~ H_2 + q_\mu^E \left( P_\mu^E \right)^2 ~ H_3 + \left( P_\mu^E \right)^3 ~ H_4 \right]\,, \\[2mm]
\label{eq:scalar_hypercubic}
&&\braket{S}_{\rm hyp} = a^2 \left[ q^{[4]} ~ \widetilde{H}_1+ q^{[3]} P^{[1]} ~ \widetilde{H}_2 + 
q^{[2]} P^{[2]} ~ \widetilde{H}_3 + q^{[1]} P^{[3]} ~ \widetilde{H}_4 + P^{[4]} \widetilde{H}_5\right]/(\mu_c - \mu_q)\,,
\end{eqnarray}
where the quantities $H_i$ ($i = 1, ..., 4$) and $\widetilde{H}_j$ ($j = 1,...,5$) are new hypercubic form factors.
Eqs.~(\ref{eq:vector_hypercubic}) and (\ref{eq:scalar_hypercubic}), which are built in terms of the two momenta $q_\mu^E$ and $P_\mu^E=(p_D+p_P)_\mu^E$, are the most general structure, up to order $\mathcal{O}(a^2)$, that transform under hypercubic rotations respectively as a four-vector and a scalar. Moreover, by studying the Ward-Takahashi Identity (WTI) relating the 4-divergence of the vector current to the scalar density (see Ref.~\cite{Lubicz:2017syv}), $\braket{S}_{\rm hyp}$ can be further simplified as:
\begin{equation}
\braket{S}_{\rm hyp} = \left[ a^2 q^{[4]} ~ H_S -  q_\mu^E \braket{\widehat{V}_\mu^E}_{\rm hyp} \right]/(\mu_c - \mu_q)\,.
\label{eq:scalar_hypercubic_final}
\end{equation}
For the form factors $H_i$ and $H_S$ we adopted the simple polynomial expressions $H_i(z) = d_0^i + d_1^i z + d_2^i z^2$ and $H_S = d_0^S + d_1^S m_\ell$, which are given in terms of the $z$ variable \cite{Boyd:1995cf,Arnesen:2005ez} and the light-quark mass $m_\ell$. The coefficients $d_{0,1,2}^{\,i}$ and $d_{0,1}^{\,S}$ are free parameters.

As for the Lorentz-invariant terms $\braket{\widehat{V}_\mu^E}_{\rm Lor}$ and $\braket{S}_{\rm Lor}$, the form factors $f_{+,0}(q^2, a^2)$ can be parametrized by the modified z-expansion of Ref.~\cite{Bourrely:2008za}, viz.
\begin{eqnarray}
\label{eq:z-exp_f+}
f_{+,0}^{D \to \pi(K)}(q^2, a^2) \!\!\!& = &\!\!\! \left[f^{D \to \pi(K)}(0, a^2) + c^{D \to \pi(K)}_{+,0}(a^2)\, (z - z_0)
	\left(1 + \frac{z + z_0}{2} \right)\right]/\mathcal{P}^{D \to \pi(K)}_{+,0}(q^2) \,,
\end{eqnarray}
where in the case of the $D \to \pi$ transition we considered the single-pole expressions
\begin{equation}
\label{eq:pole_Dpi}
\mathcal{P}_+^{D\to\pi}(q^2) = 1 - q^2 / M_V^2 \,,\quad \mathcal{P}_0^{D\to\pi}(q^2) = 1 - K_{FSE}^0(L) q^2 / M_S^2 \,,
\end{equation}
while for the $D \to K$ channel we used
\begin{equation}
\label{eq:pole_DK}
\mathcal{P}_+^{D\to K}(q^2) =  1 - q^2  \left( 1 + P_+ a^2 \right) / M_{D_s^*}^2 \,,\quad\mathcal{P}_0^{D\to K}(q^2) =  1 \,.
\end{equation}
In Eq.~(\ref{eq:z-exp_f+}) the term proportional to ($z^2 - z_0^2$) is constrained by the requirement of analyticity at the annihilation threshold~\cite{Bourrely:2008za}, and it is applicable for the vector form factor only.
However, the fitting procedure turns out to be almost insensitive to the presence of such a constraint.
The quantities $M_V$ and $M_S$ represent the vector and scalar pole masses, respectively, and they are free parameters in the fitting procedure.
On the other hand for the $D \to K\ell\nu$ process, the physical vector meson $D_s^*$ has a mass below the cut threshold $\sqrt{t_+} = (M_{D_s} + M_K)$, so the pole factor $1/M_{D_s^*}^2$, including a simple discretization effect proportional to $a^2$, is introduced to guarantee the applicability of the $z$-expansion.
Conversely, the data for the scalar form factor $f_0^{D\to K}$ can be fitted equally well both including and  excluding the pole term. For this reason we have finally set $\mathcal{P}_0^{D \to K}(q^2) = 1$.
The quantity $K_{FSE}^0(L)$ in the r.h.s. of Eq.~(\ref{eq:pole_Dpi}) takes into account the FSE, which has been observed only in the slope of the scalar form factor $f_0^{D\to \pi}$~\cite{Lubicz:2017syv}, through the following phenomenological form
\begin{equation}
K_{FSE}^0(L) = 1 + \left[C_{FSE}^{0} ~ \xi_\ell \, e^{-M_\pi L}\right]/(M_\pi L) ~,
\label{eq:FSE}
\end{equation}
where $C_{FSE}^0$ is a free parameter and $\xi_\ell = 2B m_\ell/ (16\pi^2f^2)$, with $B$ and $f$ being the SU(2) low-energy constants entering the LO chiral Lagrangian, determined in Ref.~\cite{Carrasco:2014cwa}. 
For the form factors at zero 4-momentum transfer we imposed the condition $f^{D \to \pi(K)}_+(0, a^2)=f^{D \to \pi(K)}_0(0, a^2)\equiv f^{D \to \pi(K)}(0, a^2)$ and we used the following expression:
\begin{equation}
\label{eq:ChLim}
f^{D \to \pi(K)}(0, a^2) = F_+ \left[ 1 + A^{\pi(K)}\, \xi_\ell \log\xi_\ell + b_1\, \xi_\ell + b_2 \, \xi^2_\ell + D \, a^2 \right] ~,
\end{equation}
where $F_+$, $b_1$, $b_2$ and $D$ are free parameters in the fitting procedure, while $A^{\pi}$ and $A^{K}$ are the chiral-log coefficients predicted by the hard pion SU(2) Chiral Perturbation Theory (ChPT) \cite{Bijnens:2010jg}, given by $ A^\pi =- (3/4) \left( 1 + 3\, \widehat{g}^{\,2} \right)\,$ (with $\widehat{g} = 0.61$ \cite{Olive:2016xmw}) and $A^K=1/2$.

In the limit where the parent and the child mesons are the same, Eq.~(\ref{eq:vector_hypercubic}) reduces to
\begin{equation}
\braket{D(p^\prime) | \widehat{V}_\mu^E |D(p)}_{\rm hyp} = a^2 \left[  \left( q_\mu^E \right)^2 P_\mu^E H_2 + \left( P_\mu^E \right)^3 H_4 \right] ~ .
\label{eq:vector_hypercubic_DD}
\end{equation}
Since there is no evidence of hypercubic effects when the initial and final meson have similar masses (see Fig.~\ref{fig:DtoD}), the hypercubic form factors $H_2$ and $H_4$ might be neglected. 
For this reason we have repeated the global fitting procedure assuming $H_2 = H_4 = 0$ . 
Differences in the results were found to be negligible for both the $D\to\pi$ and $D\to K$ channels, either when we included ($H_2 \neq 0$, $H_4 \neq 0$) or excluded ($H_2 = H_4 = 0$) the hypercubic form factors $H_2$ and $H_4$.
Therefore, we adopted $H_2 = H_4 = 0$ as our reference choice in the global fit for estimating uncertainties due to systematic errors as well as for obtaining the form factors $f_{+,0}^{D \to \pi(K)}(q^2)$.

The novelty of our analysis with respect to previous studies of the $D \to \pi(K)\ell\nu$ form factors, is the use of many kinematical conditions in which both the parent and the child mesons can be either in motion or at rest. The main point is that using only a limited number of kinematical conditions, like for instance the Breit-frame ($\vec{p_D} = - \vec{p}_{\pi(K)}$) or the $D-$meson at rest frame, hypercubic effects may not be correctly observed even if they are present in the data.
\begin{figure}[htb]
	\centering
		\includegraphics[width=6.7cm,clip]{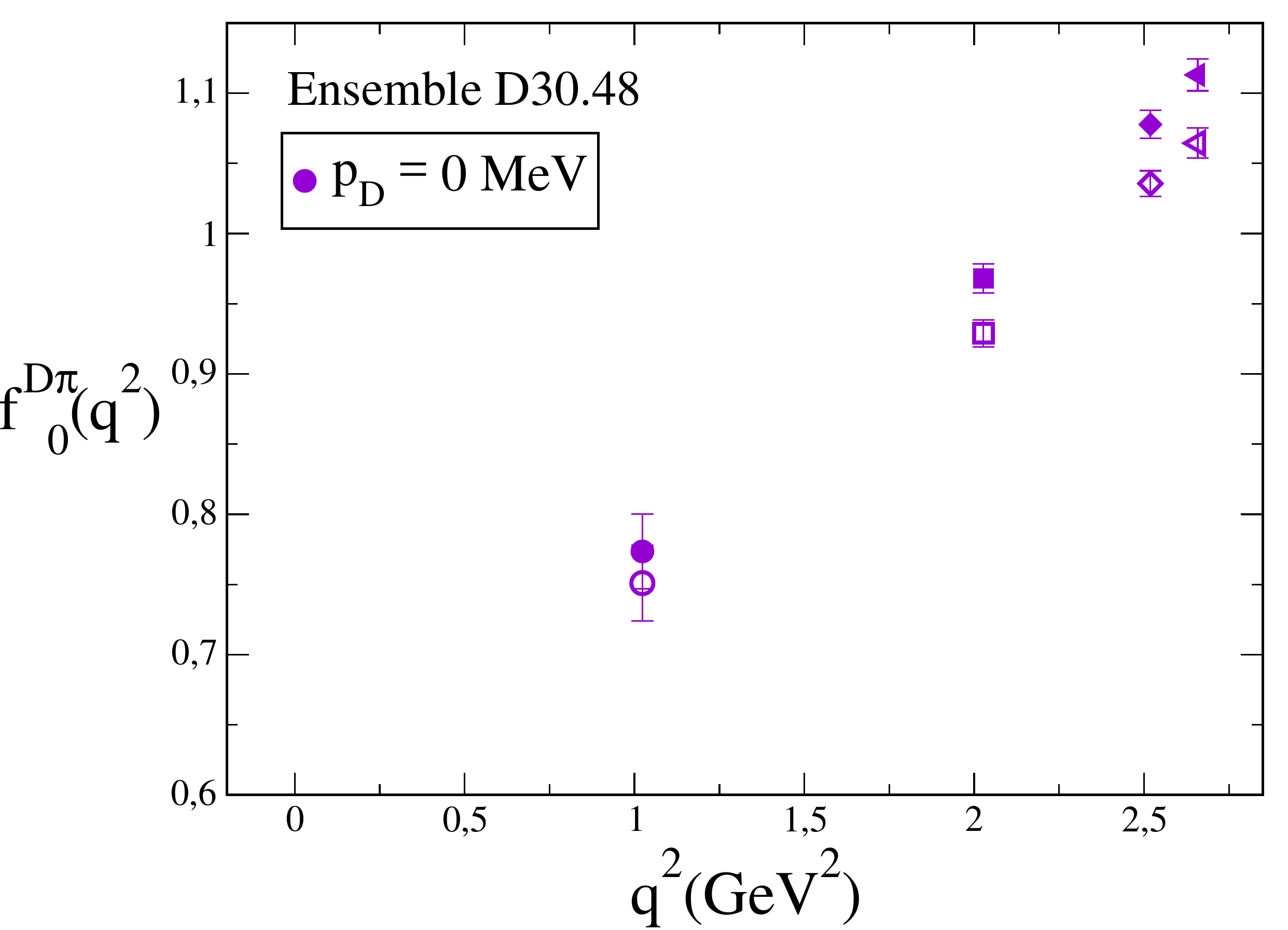} 
		\includegraphics[width=6.7cm,clip]{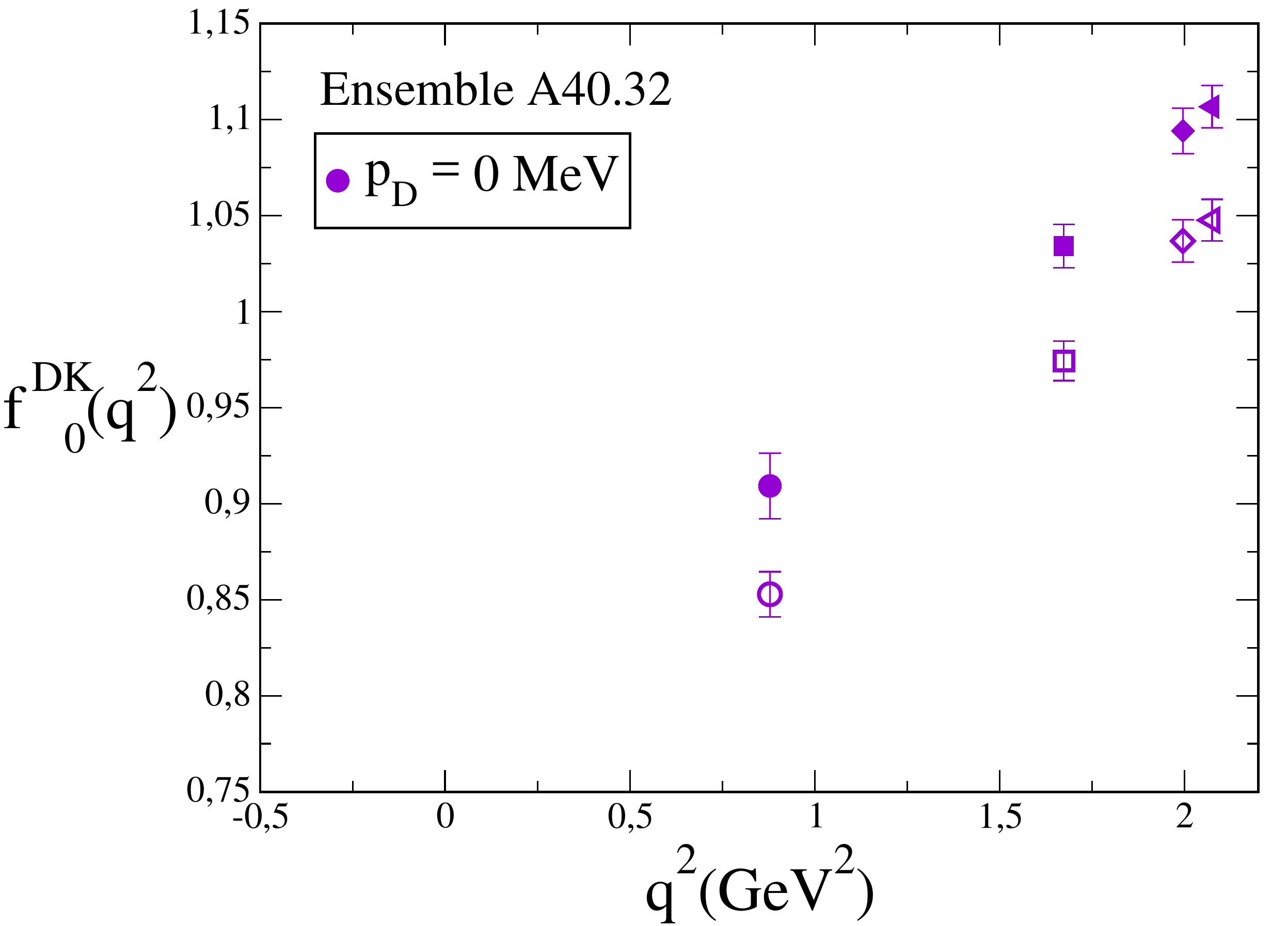}
	\caption{\it The scalar form factors $f_0^{D \pi}(q^2)$ (ensemble D30.48) and $f_0^{D K}(q^2)$ (ensemble A40.32) corresponding to the kinematical conditions with the $D-$meson at rest. Hollow and filled points represent, respectively, data before and after the removal of the hypercubic effects determined in the global fitting procedure.}
	\label{fig:D_at_rest}
\end{figure}
This is illustrated in Fig.~\ref{fig:D_at_rest}, which shows the subset of our data for the scalar form factors $f_0^{D\to\pi(K)}$ corresponding only to the D-meson at rest both before and after the subtraction of the hypercubic effects determined in the global fitting procedure.
Lorentz-symmetry breaking is not manifest in the limited set of data points with $\vec{p}_D = 0$, although its impact is not negligible.
This holds for the scalar form factor $f_0$, while in the case of the vector form factor $f_+$ we find that Lorentz-symmetry breaking effects are less pronounced in the subset of data corresponding to the D-meson at rest. 

\section{Results}

\begin{figure}[htb]
	\centering
		\includegraphics[width=7.0cm,clip]{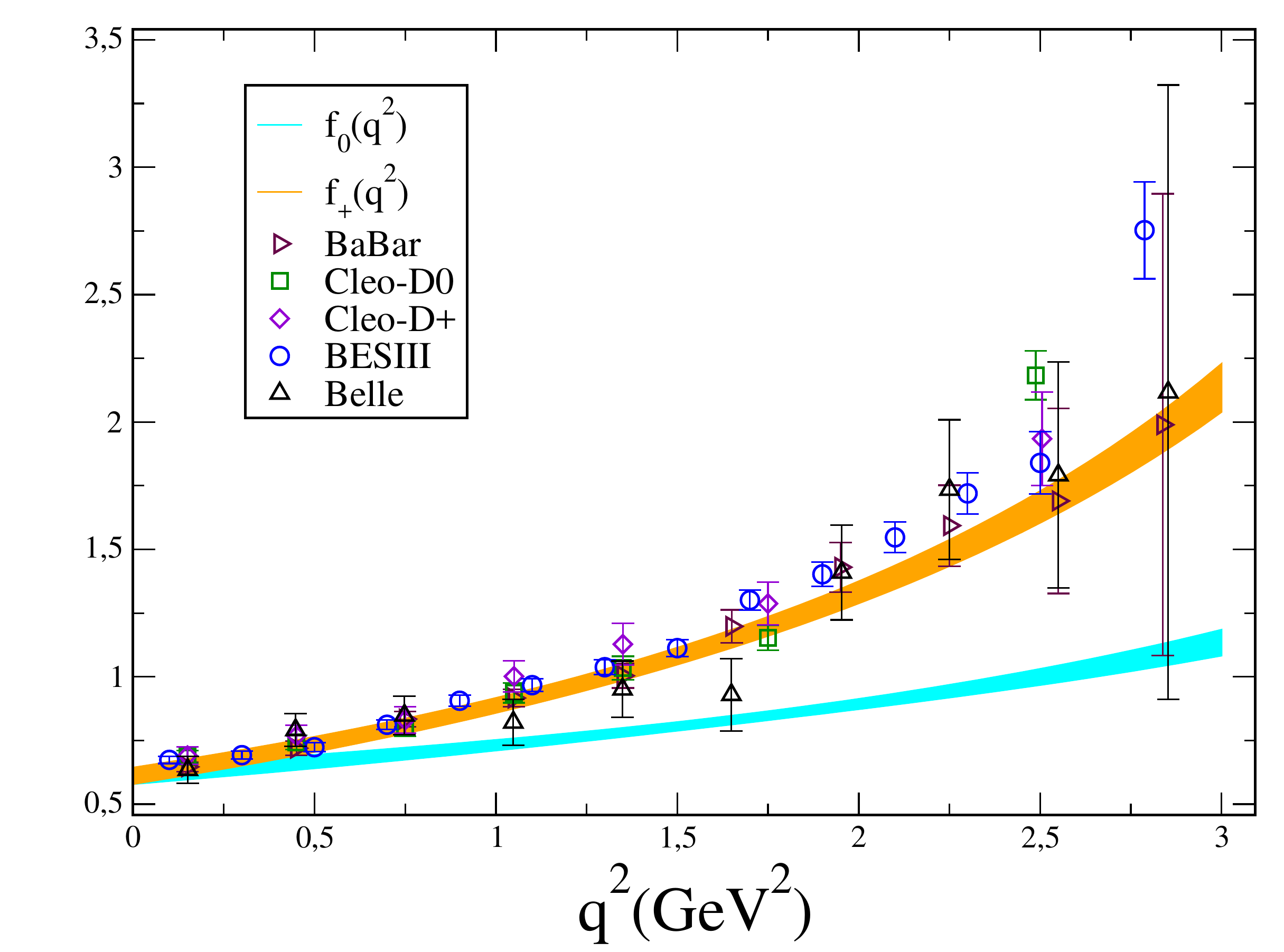} 
		\includegraphics[width=7.0cm,clip]{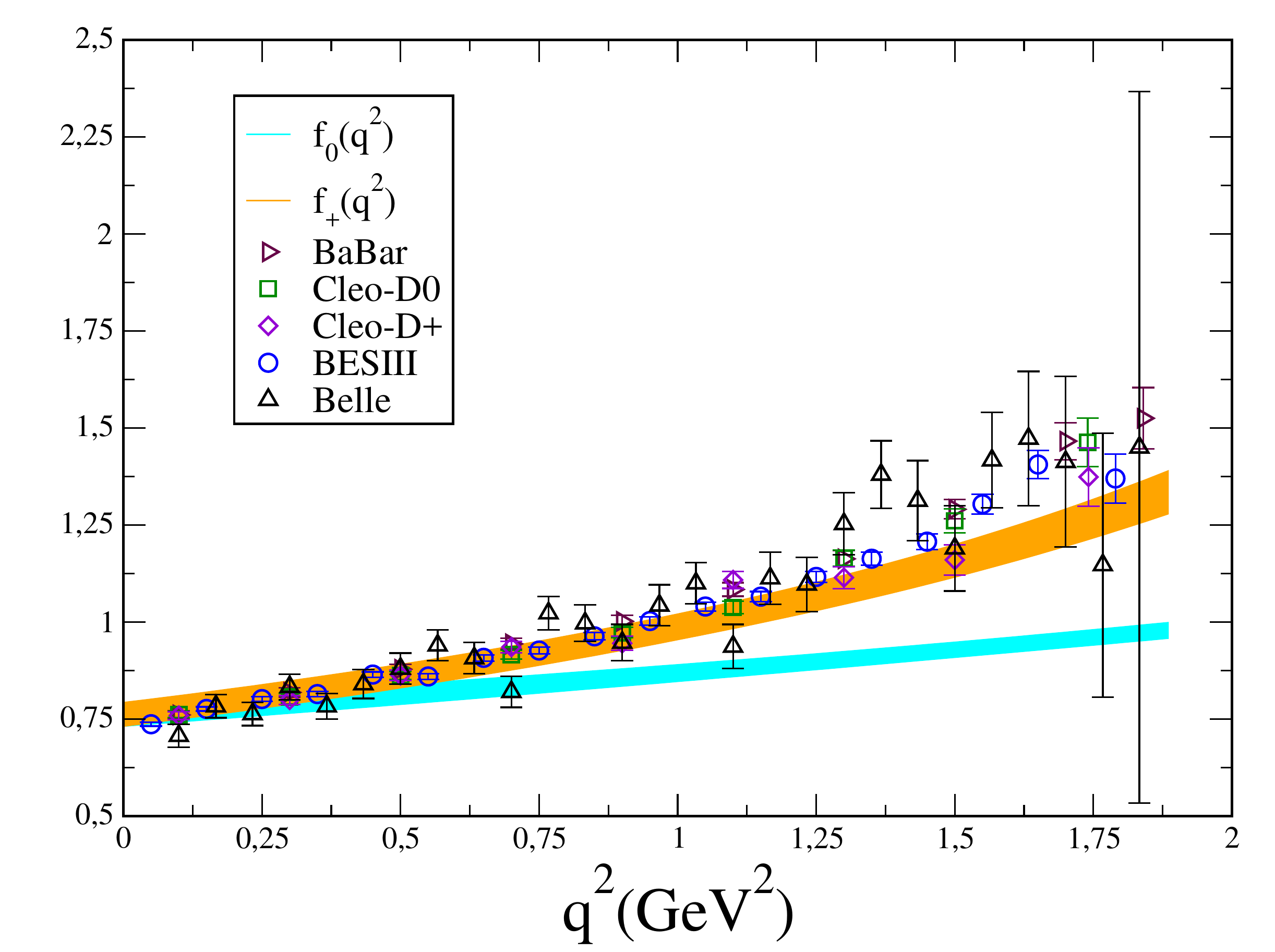}
	\caption{\it Lorentz-invariant form factors $f_+(q^2)$ (orange bands) and $f_0(q^2)$ (cyan bands), extrapolated to the physical pion mass and to the continuum and infinite volume limits, for the $D \to \pi$ (left panel) and $D \to K$ (right panel) transitions, including their total uncertainties. The values of $f^{D\pi(K)}_+(q^2)$ determined by BELLE, BABAR, CLEO and BESIII collaborations in Refs.~\cite{Widhalm:2006wz,Lees:2014ihu,Aubert:2007wg,Besson:2009uv,Ablikim:2015ixa} are shown. The bands correspond to the total (statistical + systematic) uncertainty at one standard-deviation level.}
	\label{fig:physicalFormFactors}
\end{figure}

\begin{SCfigure}[1.0][htb]
\centering
\includegraphics[width=9cm]{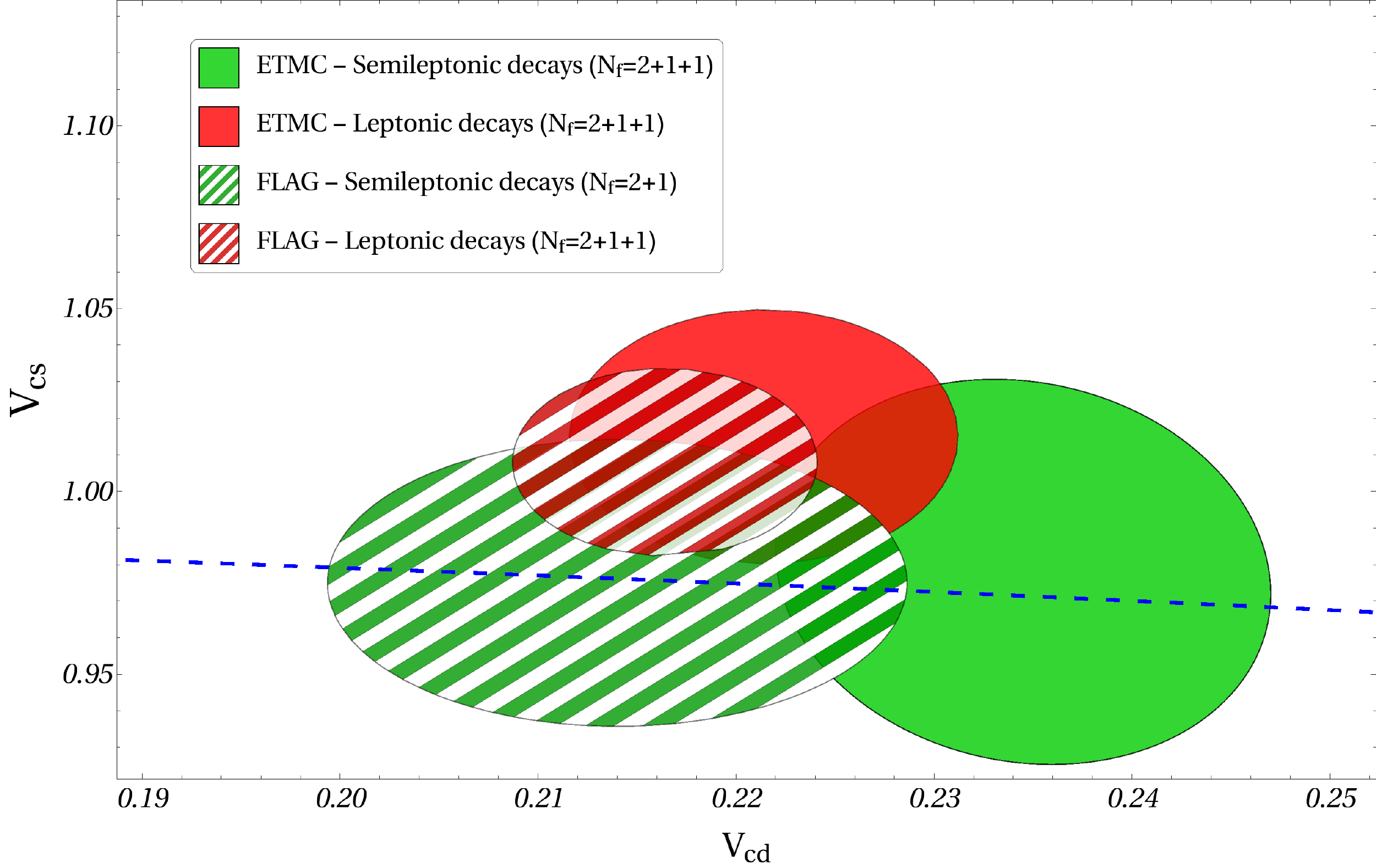} $~~~~$ \vspace{0.0mm}
\caption{\it Results for $|V_{cd}|$ and $|V_{cs}|$ obtained from leptonic and semileptonic $D$- and $D_s$-meson decays, represented respectively by green and red ellipses corresponding to a $68 \%$ probability contour. The solid ellipses are the results of Refs.~\cite{Riggio:2017zwh} and \cite{Carrasco:2014poa}. Striped ellipses correspond to the FLAG results~\cite{Aoki:2016frl}, which are based on the LQCD results obtained in Refs.~\cite{Na:2011mc,Na:2010uf} with $N_f = 2 + 1$ dynamical quarks. The dashed line indicates the $|V_{cd}| - |V_{cs}|$ correlation that follows from the CKM unitarity.}
\label{fig:VcdVcs}
\end{SCfigure}

The physical Lorentz-invariant vector and scalar form factors, extrapolated to the physical pion mass and to the continuum and infinite volume limits, are shown in Fig.~\ref{fig:physicalFormFactors} for the $D \to \pi$ and $D \to K$ transitions.
Our results are compared with the corresponding values determined by BELLE, BABAR, CLEO and BESIII collaborations in Refs.~\cite{Widhalm:2006wz,Lees:2014ihu,Aubert:2007wg,Besson:2009uv,Ablikim:2015ixa}.
The agreement is good except for high values of $q^2$, where some deviations are visible. 
At zero 4-momentum transfer we find
\begin{equation}
f_+^{D \to \pi}(0) = 0.612 ~ (35) \qquad ~ , ~ \qquad f_+^{D \to K}(0) = 0.765  ~ (31) ~ ,
\end{equation}
which are consistent with the FLAG~\cite{Aoki:2016frl} averages $f_+^{D \to \pi}(0) = 0.666 ~ (29)$, based on the result of Ref.~\cite{Na:2011mc}, and $f_+^{D \to K}(0) = 0.747 ~ (19)$ from Ref.~\cite{Na:2010uf}. 
The knowledge of the form factors in the full kinematical range allowed us to perform the first determination of the CKM matrix elements $|V_{cd}|$ and $|V_{cs}|$ in a truly consistent way within the SM, by combining directly the momentum dependence of the vector form factors $f_+^{D\to\pi(K)}(q^2)$ with the experimental determinations of the decay rates for the $D\to\pi(K)\ell\nu$ processes, without making use of any other assumption. 
The extraction of $|V_{cd}|$ and $|V_{cs}|$ have been carried out in Ref.~\cite{Riggio:2017zwh}, where we obtain:
\begin{equation}
  |V_{cd}| = 0.2345 ~ (83) \qquad ~ , ~ \qquad |V_{cs}| = 0.978 ~ (35) ~ .
\label{eq:resultsCKM_ff}
\end{equation}
Using $|V_{cb}| = 0.0360\,(9)$ from Ref.~\cite{Olive:2016xmw}, the unitarity of the second row of the CKM matrix is given by $|V_{cd}|^2 + |V_{cs}|^2 + |V_{cb}|^2 = 1.013 ~ (68)$.
The results (\ref{eq:resultsCKM_ff}) are presented in Fig.~\ref{fig:VcdVcs} as ellipses in the $|V_{cd}| - |V_{cs}|$ plane corresponding to a $68 \%$ probability contour. 
The ellipses corresponding to the leptonic and semileptonic FLAG averages~\cite{Aoki:2016frl} for $|V_{cd}|$ and $|V_{cs}|$ are also shown, as well as the constraint imposed by the second-row unitarity, indicated by a dashed line. 

\section*{Acknowledgements}
We warmly thank N.~Carrasco and F.~Sanfilippo for their valuable contribution to the initial stages of the present work.
We thank our colleagues of the ETMC for fruitful discussions.
We gratefully acknowledge the CPU time provided by PRACE under the project PRA067 and by CINECA under the initiative INFN-LQCD123 on the BG/Q system Fermi at CINECA (Italy).

\newpage

\bibliography{rifbiblio}

\end{document}

%% file: definitions.tex
\newcommand{\be}{\begin{equation}}
\newcommand{\ee}{\end{equation}}
\newcommand{\bea}{\begin{eqnarray}}
\newcommand{\eea}{\end{eqnarray}}
\newcommand{\bi}{\begin{itemize}}
\newcommand{\ei}{\end{itemize}}

\newcommand{\MeV}{\,\mathrm{MeV}}

\newcommand{\fm}{\,\mathrm{fm}}

\def\mev{{\rm MeV}}
\def\gev{{\rm GeV}}
\def\tev{{\rm TeV}}
\def\fm{{\rm fm}}
 
\def\fm{\mathrm{fm}}
\def\ev{\mathrm{e\kern-0.1em V}}
\def\kev{\mathrm{ke\kern-0.1em V}}
\def\mev{\mathrm{Me\kern-0.1em V}}
\def\gev{\mathrm{Ge\kern-0.1em V}}
\def\tev{\mathrm{Te\kern-0.1em V}}